      [1999/12/01 v1.4c Il Nuovo Cimento]
\documentclass{cimento}

\usepackage{epsfig}
\usepackage{amsmath}
\usepackage{color}
\usepackage{latexsym}
\usepackage{amssymb}
\usepackage{dsfont}

\newcommand{\barpsi}{\overline{\psi}}
\newcommand{\ud}{\mathrm{d}}

\title{Canonical and kinetic decompositions of the proton spin}
\author{C.~Lorc\'{e}\from{ins:Orsay}}
\instlist{\inst{ins:Orsay} IPNO, Universit\'e Paris-Sud, CNRS/IN2P3, 91406 Orsay, France\\
and LPT, Universit\'e Paris-Sud, CNRS, 91406 Orsay, France
}
\begin{document}

\maketitle

\begin{abstract}
We propose a short summary of the present situation concerning the proton spin decomposition. We briefly discuss some of the main controversies about the issues of gauge invariance, uniqueness and measurability. As a conclusion, we argue that part of the controversies is actually undecidable. 
\end{abstract}

\section{Introduction}

The last few years have seen numerous developments concerning the proper definition of quark and gluon contributions to the proton spin. In particular, the status and the physical relevance of the canonical angular momentum operators has been clarified thanks to the notion of gauge-invariant extensions. This allows one to render gauge invariant the interpretation of $\Delta g$ as the gluon spin contribution. Moreover, it has been shown that one can access the canonical orbital angular momentum provided that one is able to extract experimentally either the Wigner distributions or particular twist-3 distributions. For a recent review of the discussions, see Ref.~\cite{Lorce:2012rr}. 

In this short letter, we summarize the recent developments about the proton spin decomposition and briefly discuss the issues of gauge invariance, uniqueness and measurability. In section \ref{sec2}, we present the suggestion made by Chen \emph{et al.} to separate the gauge field into pure-gauge and physical terms. Although gauge invariant, this approach is not unique owing to the Stueckelberg symmetry which reflects the freedom in defining what is exactly meant by pure-gauge and physical contributions. In section \ref{sec3}, we recall the kinetic and canonical gauge-invariant definitions of quark orbital angular momentum (OAM) and argue that there exist actually infinitely many inequivalent canonical OAM operators, raising the question of deciding which is the physical one. In section \ref{sec4}, we show that the Wigner operator gives access to both the kinetic and canonical angular momentum operators, provided that one uses the appropriate Wilson lines. Finally, we conclude this letter with section \ref{sec5}.

\section{Chen \emph{et al.} decomposition and Stueckelberg symmetry}\label{sec2}

In order to unambiguously define what is meant by gluon spin and orbital angular momentum, Chen \emph{et al.} proposed to separate explicitly the gauge degrees of freedom from the physical ones \cite{Chen:2008ag,Chen:2011zzh,Wakamatsu:2010qj,Wakamatsu:2010cb}
\begin{equation}\label{Adecomp}
A_\mu(x)=A^\text{pure}_\mu(x)+A^\text{phys}_\mu(x),
\end{equation}
where the pure-gauge and physical parts satisfy specific gauge transformation laws
\begin{align}
A_\mu^\text{pure}(x)&\mapsto\tilde A_\mu^\text{pure}(x)=U(x)\left[A_\mu^\text{pure}(x)+\frac{i}{g}\,\partial_\mu\right]U^{-1}(x),\label{Apureg}\\
A_\mu^\text{phys}(x)&\mapsto\tilde A_\mu^\text{phys}(x)=U(x)A_\mu^\text{phys}(x)U^{-1}(x).\label{Aphysg}
\end{align}
Since $A^\text{pure}_\mu(x)$ is a pure gauge, it can be written as
\begin{equation}
A_\mu^\text{pure}(x)=\frac{i}{g}\,U_\text{pure}(x)\partial_\mu U^{-1}_\text{pure}(x),
\end{equation}
where $U_\text{pure}(x)$ is some unitary gauge matrix  with the gauge transformation law
\begin{equation}\label{Ugauge}
U_\text{pure}(x)\mapsto\tilde U_\text{pure}(x)=U(x)U_\text{pure}(x).
\end{equation}
Clearly, in the gauge $U(x)=U^{-1}_\text{pure}(x)$ the pure-gauge term vanishes.

By construction, the decomposition \eqref{Adecomp} is gauge invariant $\tilde A_\mu=\tilde A_\mu^\text{pure}+\tilde A_\mu^\text{phys}$. However, it is not unique since we still have some freedom in defining exactly what we mean by `pure-gauge' and `physical'. The reason is that the pure-gauge and physical terms remain respectively pure-gauge and physical under the following transformation leaving $A_\mu(x)$ invariant
\begin{align}
A^\text{pure}_\mu(x)&\mapsto A^{\text{pure},g}_\mu(x)=A^\text{pure}_\mu(x)+\frac{i}{g}\,U_\text{pure}(x)U_0^{-1}(x)\left[\partial_\mu U_0(x)\right]U^{-1}_\text{pure}(x),\label{stueck1}\\
A^\text{phys}_\mu(x)&\mapsto A^{\text{phys},g}_\mu(x)=A^\text{phys}_\mu(x)-\frac{i}{g}\,U_\text{pure}(x)U_0^{-1}(x)\left[\partial_\mu U_0(x)\right]U^{-1}_\text{pure}(x),\label{stueck2}
\end{align}
where $U_0(x)$ is a gauge-invariant unitary matrix. At the level of $U_\text{pure}(x)$, this transformation reads
\begin{equation}\label{Ustueck}
U_\text{pure}(x)\mapsto U^g_\text{pure}(x)=U_\text{pure}(x)U^{-1}_0(x).
\end{equation}
While the ordinary gauge transformation acts on the left of $U_\text{pure}(x)$ as in Eq. \eqref{Ugauge}, this new transformation acts on the right. It is therefore important to distinguish them. Noting that the pure-gauge term $A^\text{pure}_\mu$ plays a role similar to the derivative of the Stueckelberg field, we refer to this transformation as the Stueckelberg (gauge) transformation \cite{Lorce:2012rr}. Explicit realizations of the Chen \emph{et al.} decomposition are usually non-local. Gauge invariance is then assured by the use of Wilson lines whose path dependence is at the origin of the Stueckelberg symmetry \cite{Lorce:2012ce}.

\section{Kinetic and canonical orbital angular momentum}\label{sec3}

There exist essentially two kinds of gauge-invariant quark orbital angular momentum. One is the \emph{kinetic} OAM \cite{Ji:1996ek}
\begin{equation}
\mathcal M^{\mu\nu\rho}_{q,\text{OAM}}(x)=\frac{i}{2}\,\barpsi(x) \gamma^\mu x^{[\nu}\!\!\stackrel{\leftrightarrow}{D}\!\!\!\!\!\phantom{\partial}^{\rho]}(x)\psi(x)
\end{equation}
and the other one is the \emph{canonical} OAM \cite{Chen:2008ag,Chen:2011zzh}
\begin{equation}\label{canonicalOAM}
\mathsf M^{\mu\nu\rho}_{q,\text{OAM}}(x)=\frac{i}{2}\,\barpsi(x) \gamma^\mu x^{[\nu}\!\!\stackrel{\leftrightarrow}{D}\!\!\!\!\!\phantom{\partial}^{\rho]}_\text{pure}(x)\psi(x),
\end{equation}
where the covariant derivatives at the point $x$ are defined as $D^\mu(x)=\partial^\mu-igA^\mu(x)$ and $D^\mu_\text{pure}(x)=\partial^\mu-igA^\mu_\text{pure}(x)$.  We used for convenience the notations $a^{[\mu}b^{\nu]}=a^\mu b^\nu-a^\nu b^\mu$ and $\stackrel{\leftrightarrow}{\partial}\,=\,\stackrel{\rightarrow}{\partial}-\stackrel{\leftarrow}{\partial}$. These two OAMs differ by a so-called potential term \cite{Wakamatsu:2010qj,Wakamatsu:2010cb}
\begin{equation}
\mathsf M^{\mu\nu\rho}_\text{pot}(x)=-g\,\barpsi(x) \gamma^\mu x^{[\nu}A^{\rho]}_\text{phys}(x)\psi(x),
\end{equation}
which is usually non-vanishing. In the gauge $U(x)=U^{-1}_\text{pure}(x)$, the canonical OAM simply reduces to the same expression as in the definition of the Jaffe-Manohar OAM \cite{Jaffe:1989jz} and can then be thought of as a gauge-invariant extension (GIE) of the latter \cite{Ji:2012sj,Ji:2012gc,Ji:2012ba}.

Contrary to the kinetic quark OAM, the canonical quark OAM is not Stueckelberg invariant, \emph{i.e.} it depends on how one explicitly separates the gauge field into pure-gauge and physical terms. There is consequently an infinite number of possible different definitions of canonical OAM, all sharing the same formal structure \eqref{canonicalOAM}. The reduction to the Jaffe-Manohar OAM occurs in different gauges, which implies that the different canonical OAMs are not equivalent. This raises the question of deciding which canonical decomposition is the physical one.

\section{Wigner operator and its relation with orbital angular momentum}\label{sec4}

The gauge-invariant quark Wigner operator is defined as \cite{Ji:2003ak,Belitsky:2003nz}
\begin{equation}\label{OAMWigner}
W^{[\gamma^\mu]q}(x,k)\equiv\int\frac{\ud^4z}{(2\pi)^4}\,e^{ik\cdot z}\,\barpsi(x-\tfrac{z}{2}) \gamma^\mu\,\mathcal W_{\mathcal C}(x-\tfrac{z}{2},x+\tfrac{z}{2})\,\psi(x+\tfrac{z}{2}).
\end{equation}
It can be interpreted as a phase-space density operator. It then is natural to define the quark OAM density as \cite{Lorce:2011kd,Lorce:2011ni}
\begin{equation}\label{OAMformula}
M^{\mu\nu\rho}_{q,\text{OAM}}(x)=\int\ud^4k\,x^{[\nu}k^{\rho]}\,W^{[\gamma^\mu]q}(x,k).
\end{equation}

In order to be gauge invariant, the definition of the Wigner operator involves a gauge link. The consequence of this gauge link is that the Wigner distribution inherits a path dependence. Using a straight gauge link in Eq.~\eqref{OAMWigner} leads to the \emph{kinetic} OAM $L_z$ \cite{Lorce:2012ce,Ji:2012sj}. With the view of connecting the Wigner distributions to the Transverse-Momentum dependent parton  Distributions (TMDs) \cite{Meissner:2009ww,Lorce:2011dv} appearing in the description of high-energy semi-inclusive processes like Semi-Inclusive DIS and Drell-Yan, it is more natural to consider instead a staple-like gauge link consisting of two longitudinal straight lines connected at $x^-=\pm\infty$ by a transverse straight line. In this case, Eq.~\eqref{OAMWigner} gives the \emph{canonical} OAM $\ell_z$ appearing in the light-front GIE \cite{Lorce:2012ce,Ji:2012sj,Lorce:2011ni,Hatta:2011ku}. 

As emphasized in the previous section, there exist formally an infinity of gauge-invariant canonical quark OAM. Note however that the proton structure is usually probed in high-energy scattering experiments. Even though physics is invariant under rotations, actual high-energy experiments provide us with a specific direction and make therefore the light-front GIE more natural, just like a Stern-Gerlach experiment provides us with a natural basis for describing the spin states.

\section{Conclusion}\label{sec5}

Separating explicitly the gauge degrees of freedom from the physical ones led to the notion of gauge-invariant extension, and allowed the definition of gauge-invariant canonical angular momentum operators. This approach has been shown to be tightly connected with the use of non-local operators and Wilson lines. However, the gauge-invariant canonical operators are not unique owing to the Stueckelberg symmetry that can be thought of as the path dependence in the non-local approach. This is nicely reflected in the definition of quark orbital angular momentum based on the Wigner operators. Depending on the choice of the path for the Wilson line, one obtains different gauge-invariant results. We stress that it is the experimental conditions that fix the Wilson lines and therefore the gauge-invariant extension to use.

\acknowledgments

I would like to thank E.~Leader, A. Metz, B. Pasquini, L. Szymanowski, M.~Wakamatsu and F.~Wang for many helpful comments and discussions. This work was supported by the P2I (``Physique des deux Infinis'') network.


\begin{thebibliography}{0}

\bibitem{Lorce:2012rr} 
  C.~Lorc\'e, to appear in PRD (2013),
  %``Geometrical approach to the proton spin decomposition,''
  arXiv:1205.6483 [hep-ph].
  %%CITATION = ARXIV:1205.6483;%%

\bibitem{Chen:2008ag} 
  X.~-S.~Chen, X.~-F.~Lu, W.~-M.~Sun, F.~Wang and T.~Goldman,
  %``Spin and orbital angular momentum in gauge theories: Nucleon spin structure and multipole radiation revisited,''
  Phys.\ Rev.\ Lett.\  {\bf 100}, 232002 (2008).
  %[arXiv:0806.3166 [hep-ph]].
  %%CITATION = ARXIV:0806.3166;%%

\bibitem{Chen:2011zzh} 
  X.~-S.~Chen, W.~-M.~Sun, F.~Wang and T.~Goldman,
  %``Art of spin decomposition,''
  Phys.\ Rev.\ D {\bf 83}, 071901 (2011).
  %[arXiv:1105.6304 [hep-ph]].
  %%CITATION = ARXIV:1105.6304;%%

\bibitem{Wakamatsu:2010qj} 
  M.~Wakamatsu,
  %``On Gauge-Invariant Decomposition of Nucleon Spin,''
  Phys.\ Rev.\ D {\bf 81}, 114010 (2010).
  %[arXiv:1004.0268 [hep-ph]].
  %%CITATION = ARXIV:1004.0268;%%

\bibitem{Wakamatsu:2010cb} 
  M.~Wakamatsu,
  %``Gauge and frame-independent decomposition of nucleon spin,''
  Phys.\ Rev.\ D {\bf 83}, 014012 (2011).
 %[arXiv:1007.5355 [hep-ph]].
  %%CITATION = ARXIV:1007.5355;%%

\bibitem{Lorce:2012ce} 
  C.~Lorc\'e, to appear in PLB (2013), 
  %``Wilson lines and orbital angular momentum,''
  arXiv:1210.2581 [hep-ph].
  %%CITATION = ARXIV:1210.2581;%%

\bibitem{Ji:1996ek} 
  X.~-D.~Ji,
  %``Gauge invariant decomposition of nucleon spin and its spin - off,''
  Phys.\ Rev.\ Lett.\  {\bf 78}, 610 (1997).
  %[hep-ph/9603249].
  %%CITATION = HEP-PH/9603249;%%

\bibitem{Jaffe:1989jz} 
  R.~L.~Jaffe and A.~Manohar,
  %``The G(1) Problem: Fact and Fantasy on the Spin of the Proton,''
  Nucl.\ Phys.\ B {\bf 337}, 509 (1990).
  %%CITATION = NUPHA,B337,509;%%

\bibitem{Ji:2012sj} 
  X.~Ji, X.~Xiong and F.~Yuan,
  %``Proton Spin Structure from Measurable Parton Distributions,''
  Phys.\ Rev.\ Lett.\  {\bf 109}, 152005 (2012).
  %[arXiv:1202.2843 [hep-ph]].
  %%CITATION = ARXIV:1202.2843;%%

\bibitem{Ji:2012gc} 
  X.~Ji, Y.~Xu and Y.~Zhao,
  %``Gluon Spin, Canonical Momentum, and Gauge Symmetry,''
  JHEP {\bf 1208}, 082 (2012).
 % [arXiv:1205.0156 [hep-ph]].
  %%CITATION = ARXIV:1205.0156;%%

\bibitem{Ji:2012ba} 
  X.~Ji, X.~Xiong and F.~Yuan,
  %``Probing Parton Orbital Angular Momentum in Longitudinally Polarized Nucleon,''
  arXiv:1207.5221 [hep-ph].
  %%CITATION = ARXIV:1207.5221;%%

\bibitem{Ji:2003ak}
  X.~Ji,
  %``Viewing the proton through 'color'-filters,''
  Phys.\ Rev.\ Lett.\  {\bf 91}, 062001 (2003).
  %[arXiv:hep-ph/0304037].
  %%CITATION = PRLTA,91,062001;%%

\bibitem{Belitsky:2003nz}
  A.~V.~Belitsky, X.~Ji and F.~Yuan,
  %``Quark imaging in the proton via quantum phase-space distributions,''
  Phys.\ Rev.\  D {\bf 69}, 074014 (2004).
  %[arXiv:hep-ph/0307383].
  %%CITATION = PHRVA,D69,074014;%%

\bibitem{Lorce:2011kd} 
  C.~Lorc\'e and B.~Pasquini,
  %``Quark Wigner Distributions and Orbital Angular Momentum,''
  Phys.\ Rev.\ D {\bf 84}, 014015 (2011).
  %[arXiv:1106.0139 [hep-ph]].
  %%CITATION = ARXIV:1106.0139;%%

\bibitem{Lorce:2011ni} 
  C.~Lorc\'e, B.~Pasquini, X.~Xiong and F.~Yuan,
  %``The quark orbital angular momentum from Wigner distributions and light-cone wave functions,''
  Phys.\ Rev.\ D {\bf 85}, 114006 (2012).
  %[arXiv:1111.4827 [hep-ph]].
  %%CITATION = ARXIV:1111.4827;%%

\bibitem{Meissner:2009ww}
  S.~Meissner, A.~Metz and M.~Schlegel,
  %``Generalized parton correlation functions for a spin-1/2 hadron,''
  JHEP {\bf 0908}, 056 (2009).
%  [arXiv:0906.5323 [hep-ph]].
  %%CITATION = JHEPA,0908,056;%%

\bibitem{Lorce:2011dv} 
  C.~Lorc\'e, B.~Pasquini and M.~Vanderhaeghen,
  %``Unified framework for generalized and transverse-momentum dependent parton distributions within a 3Q light-cone picture of the nucleon,''
  JHEP {\bf 1105}, 041 (2011).
 % [arXiv:1102.4704 [hep-ph]].
  %%CITATION = ARXIV:1102.4704;%%

\bibitem{Hatta:2011ku} 
  Y.~Hatta,
  %``Notes on the orbital angular momentum of quarks in the nucleon,''
  Phys.\ Lett.\ B {\bf 708}, 186 (2012).
  %[arXiv:1111.3547 [hep-ph]].
  %%CITATION = ARXIV:1111.3547;%%


\end{thebibliography}
\end{document}